\documentclass[aps,prl,preprint,superscriptaddress,longbibliography]{revtex4-1}
\usepackage{natbib}
\usepackage{amsmath}
\usepackage{graphicx}
\usepackage{color}
\usepackage[normalem]{ulem}

\usepackage{array}
\newcommand{\PreserveBackslash}[1]{\let\temp=\\#1\let\\=\temp}
\newcolumntype{C}[1]{>{\PreserveBackslash\centering}p{#1}}

\begin{document}

\title{Definitive Detection of Orbital Angular Momentum States in Neutrons by Spin-polarized $^3$He}

\author{Terrence Jach}
\email{terrence.jach@nist.gov}
\affiliation{Material Measurement Laboratory, National Institute of Standards and Technology, 100 Bureau Drive, Gaithersburg, MD 20899}

\author{John Vinson}
\affiliation{Material Measurement Laboratory, National Institute of Standards and Technology, 100 Bureau Drive, Gaithersburg, MD 20899}

\begin{abstract}
A standard method to detect thermal neutrons is the  nuclear interaction $^3$He(n,p)$^3$H. The spin-dependence of this interaction is also the basis of a neutron spin-polarization filter using nuclear polarized $^3$He. We consider the corresponding interaction for neutrons placed in an intrinsic orbital angular momentum (OAM) state. We derive the relative polarization-dependent absorption cross-sections for neutrons in an $L=1$ OAM state.  The absorption of those neutrons results in compound states $J^\pi=0^-$, $1^-$, and $2^-$. Varying the three available polarizations tests that an OAM neutron has been absorbed and probes which decay states are physically possible. We describe the energetically likely excited states of $^4$He after absorption, due to the fact that the compound state has odd parity. This provides a definitive method for detecting neutron OAM states and suggests that intrinsic OAM states offer the possibility to observe new physics, including anomalous cross-sections and new channels of radioactive decay.
\end{abstract}


\maketitle


Intrinsic Orbital Angular Momentum (OAM) states are quantum states in which the wave packet of a particle is given a helicity by retarding its phase progressively around its axis of travel.  The creation of observable intrinsic OAM states in photons and electrons has been convincingly observed and demonstrated \cite{Allen,Leach,Lloyd,Bliokh}. It has generated a great deal of  interest in the possible creation and observation of OAM states of thermal neutrons. Several theoretical schemes and experimental methods have been reported \cite{Clark,PRL121,NJP20,PNAS,Geerits}. While the experiments show effects compatible with neutron OAM states, they are unable to rule out non-OAM explanations of the observed data 
\cite{Cappelletti,Cappelletti2020}. The genuine creation of a neutron in an intrinsic quantum OAM state can only be demonstrated convincingly by a single-particle interaction that produces measurable quantum states. 

One of the principal channels of detecting thermal neutrons is the reaction 
\begin{equation}
\label{Interaction}
n+ {}^3\mathrm{He}=p+{}^3\mathrm{H}+764\, \mathrm{keV} \,,
\end{equation}
where the kinetic energies of the decay products are $E_p=573$~keV and $E_{{}^3\mathrm{H}}=191$~keV. The absorption cross-section is highly dependent on the angular momentum of the oriented nuclei. While the actual nuclear matrix elements (dependent only on $J$) are not easily determined, their dependence on angular momentum alignment, and therefore polarization, is readily separated out due to the Wigner-Eckart Theorem. 

Cross-sections were derived for absorption into the singlet and triplet states by Rose \cite{Rose}, predicting strong dependence of the capture of neutrons by $^3$He on their respective polarizations. The spin-dependence of oriented thermal neutrons in reaction with oriented $^3$He was first measured by Passell and Schermer \cite{Passell}, who determined that the nuclear interaction was consistent with absorption occurring exclusively in the singlet state, $J^\pi=0^+$.

In this paper we propose that the thermal capture of a spin-polarized neutron in an OAM state, absorbed by a spin-polarized $^3$He nucleus, will provide definitive proof of the neutron OAM state. We derive the dependence of the absorption on the aligned angular momentum of a neutron OAM state (with $L=1$) and that of the $^3$He. We show that the resulting angular momentum dependent cross-sections, with $J^\pi=0^-$, $1^-$, and $ 2^-$, vary with the available polarizations in a manner that cannot be obtained in the case of ordinary neutrons.  The odd parity and the energetics of the neutron absorption suggest which J states are likely to occur and what decay schemes will result.

We will discuss only the case for thermal neutrons. Thus the interaction of the neutron with the $^3$He nucleus is purely s-wave in the scattering plane, although orbital angular momentum may be added perpendicular to that plane. An important characteristic of the derivations is to express the results in terms only of the spin-polarization $p$ of the neutrons, the polarization $P_L$ of the OAM states, and the spin-polarization $P_N$ of the $^3$He  nuclei, assuming that these are the only parameters at our disposal in an experiment. 

 We note that in each case, the quantum state is determined at every step along the way. In other words, a neutron is placed in a specific spin state by a process, a specific amount of orbital angular momentum is added with a specific direction relative to the same axis by a device, and it interacts with $^3$He that has been placed into a specific oriented spin state relative to the same axis. The polarizations specify relative numbers of neutrons in specific (i.e. parallel and antiparallel) states. While it may be possible to create OAM states that are not parallel to the wavevector of the neutron \cite{Geerits}, all polarizations here are regarded as helicities along the wavevector of the neutron. 

 Here we take the original calculation of Rose as a starting point to obtain the form of cross-sections that would be observed from spin-polarized neutrons \cite{Rose}. Assume that we have a neutron with a spin wave function $\chi$ with angular momentum $S=1/2$ and $m_S=\mu=\pm1/2$. The  $^3$He nucleus has a nuclear spin wave function $\psi$ with a spin $j_N$ where $m_N=\pm1/2$. The compound nucleus in state $\Xi$ formed by the absorption of the neutron will have an angular momentum $j'=j_N{\pm}S$ with $m_{j'}=m_N+\mu$.

The cross-section depends on the distribution of spins,
\begin{equation}
\label{Rose1}
\sigma=K(j')\sum_{m_N,\mu}p(m_N)p(\mu)|\langle\Xi_{j',m'}|\psi_{j_N,m_N}\chi_{\frac{1}{2},\mu}\rangle|^2,
\end{equation}
where $K(j')$ is the squared nuclear part of the wave function that we will regard as a constant dependent only on $j'$.
The quantities $p(m_N)$ and $p(\mu)$ are the probabilities that the spin alignments are $m_N$ and $\mu$.

The probability $p(\mu)$ is generally defined by 
\begin{equation}
\label{Rose3}
p({\mu=\pm}1/2)=\frac{p_\pm}{p_++p_-}\, ,
\end{equation} 
the probability that the neutron will be parallel $(+)$ or antiparallel$(-)$ to its direction of travel.

The neutron spin polarization will then be defined in the conventional manner by 
\begin{equation}
\label{Rose4}
p=\frac{p_+-p_-}{p_++p_-}.
\end{equation}

The nuclear spin polarization is defined by \cite{Rose}

\begin{equation}
\label{Rose6}
P_N=\frac{1}{j}\sum_{m_N}{m_N}p(m_N).
\end{equation}

Thus for $^3$He, the nuclear spin probability $p(m_N)$ and spin polarization $P_N$ are given by
\begin{equation}
\label{Rose8}
p(m_N)=\frac{P_{N\pm}}{P_{N+}+P_{N-}} \; ; \quad P_N=\frac{P_{N+}-P_{N-}}{P_{N+}+p_{N-}}.
\end{equation}

The cross-section then becomes 
\begin{equation}
\label{Rose9}
\sigma=K(j')\sum_{m_N,\mu}p(m_N)p(\mu)|C(j',m'|j_N,m_N;1/2,\mu)|^2,
\end{equation}
where $\langle\Xi_{j',m'}|\psi_{j,m}\chi_{\frac{1}{2},\mu}\rangle=C(j',m'|j,m;1/2,\mu)$, the Clebsch-Gordan coefficient for $\mu={\pm}1/2$.

The cross-section (Eq.~\ref{Rose9}) can be evaluated for the triplet case $j'=j_N+\frac{1}{2}$ and the singlet case $j'=j_N-\frac{1}{2}$. 
After some algebra, the cross-section for the triplet case in terms of the polarizations becomes
\begin{equation}
\label{Rose10}
\sigma_1=\frac{K(j'=1)}{4}[3+P_Np] ,
\end{equation}
and the cross-section for the singlet case becomes
\begin{equation}
\label{Rose11}
\sigma_0=\frac{K(j'=0)}{4}[1-P_Np].
\end{equation}

If the neutrons and the $^3$He nuclei are both polarized and parallel such that, $pP_N=1$, then $\sigma_0=0$.
Measurements subsequent to that of Passell and Schermer \cite{Passell} have confirmed that the interaction of the neutron with the $^3$He nucleus is only through the singlet channel \cite{Borzakov,Zimmer}. This is the basis for using optically pumped $^3$He gas as a neutron spin polarizing filter \cite{PhysicaB}.

We now extend this to the case of an orbital angular momentum state of the neutron. We limit ourselves to the case where a device will transform the neutrons into an OAM state of $j_L=1$. Based on the definition of polarization (Eq.~\ref{Rose6}) and $m_L=\pm1$, the OAM polarization is given by
\begin{equation}
\label{OAM1}
P_L=\frac{P_{L+}-P_{L-}}{P_{L+}+P_{L-}}.
\end{equation}
We define a total neutron angular momentum state $\Phi$ made up of the neutron OAM state $\zeta_L$ and the neutron spin state $\chi_S$. 

We expect that the neutron spin and orbital angular momentum will combine to form two possible states.
For $j'=j_L+j_S=\frac{3}{2}$,
\begin{equation}
\label{OAM2}
|\Phi\rangle=\sum_{m_L,\mu}|\Phi_{\frac{3}{2},m'}{\rangle}\langle\Phi_{\frac{3}{2},m'}|\zeta_{j_L,m_L}\chi_{\frac{1}{2},\mu}\rangle,
\end{equation}
where $m'=m_L+\mu$ and $\langle\Phi_{\frac{3}{2},m'}|\zeta_{j_L,m_L}\chi_{\frac{1}{2},\mu}\rangle=C(3/2,m'|j_L,m_L;1/2,\mu)$, the Clebsch-Gordan coefficient.
For $j'=j_L-j_S=\frac{1}{2}$,
\begin{equation}
\label{OAM3}
|\Phi\rangle=\sum_{m_L,\mu}|\Phi_{\frac{1}{2},m'}{\rangle}\langle\Phi_{\frac{1}{2},m'}|\zeta_{j_L,m_L}\chi_{\frac{1}{2},\mu}\rangle,
\end{equation}
where $m'=m_L+\mu$ and $\langle\Phi_{\frac{1}{2},m'}|\zeta_{j_L,m_L}\chi_{\frac{1}{2},\mu}\rangle=C(1/2,m'|j_L,m_L;1/2,\mu)$.

Since we are only able to control $m_L$ and $\mu$ in our experiment, we end up with spin-polarized OAM neutrons in  linear combinations of states as shown in Table~1. When absorbed, we expect that the states $|\Phi\rangle$ will combine with the $^3$He nucleus and its angular momentum wave function $\psi=\psi_N$  to form a compound nucleus in a state $\Xi$.

\begin{table}[t]
\centering
\caption{Possible OAM neutron states resulting from the control of the polarizations}
\begin{tabular}{ C{12mm}  C{12mm}  C{12mm}  l}
$m'$ & $m_L$ & $\mu$ &\;\;\; {State} \\
\hline
\hline
\noalign{\vskip 2mm}    
$+\frac{3}{2}$ & $+1$ & $+\frac{1}{2}$ & $\;\;\;\;\;\;\;\;\;\,|\Phi_{\frac{3}{2},+\frac{3}{2}}\rangle$ \\

$+\frac{1}{2}$ & $+1$ & $-\frac{1}{2}$ & $\;\;\;\;\sqrt{\frac{1}{3}}|\Phi_{\frac{3}{2},+\frac{1}{2}}\rangle+\sqrt{\frac{2}{3}}|\Phi_{\frac{1}{2},+\frac{1}{2}}\rangle$ \\

$-\frac{1}{2}$ & $-1$ & $+\frac{1}{2}$ & $\;\;\;\;\sqrt{\frac{1}{3}}|\Phi_{\frac{3}{2},-\frac{1}{2}}\rangle-\sqrt{\frac{2}{3}}|\Phi_{\frac{1}{2},-\frac{1}{2}}\rangle$ \\

$-\frac{3}{2}$ & $-1$ & $-\frac{1}{2}$ &\;\;\;\;\;\;\; \;\,$|\Phi_{\frac{3}{2},-\frac{3}{2}}\rangle$ \\
\noalign{\vskip 2mm}    
\hline
\hline
\end{tabular}
\end{table}

The possible final compound states will have angular momentum $j''=0$, $1$, and $2$. The final cross-section takes the form
\begin{align}
\label{OAM4}
\sigma&=K(j'')\sum_{m_N,m_L,\mu}p(m_N)p(m_L)p(\mu) \times \\ 
&\times \left|\sum_{j'}{\langle\Xi_{j'',m''}|\Phi_{j',m'}\psi_{\frac{1}{2},m_N}\rangle}{\langle\Phi_{j',m'}|\zeta_{j_L,m_L}\chi_{\frac{1}{2},\mu}\rangle}\right|^2, \nonumber
\end{align}
where $m''=m'+m_N$.

For $j''=0$, only $j'=\frac{1}{2}$ states of $\Phi$ are involved and for $j''=2$, only $j'=\frac{3}{2}$ states, so  Eq.~\ref{OAM4} reduces simply to   
\begin{align}
\label{OAM5}
\sigma&=K(j'')\sum_{m_N,m_L,\mu}p(m_N)p(m_L)p(\mu)\times \\
&\times \left|C(j'',m''|j',m';1/2,m_N)C(j',m'|j_L,m_L;1/2,\mu)\right|^2. \nonumber
\end{align}
A final state of $j''=1$ will result in interference from the linear combination of both neutron OAM states as indicated in Table 1.

Evaluating all the coefficients and substituting polarizations for the probabilities,
we get the final expressions for the relative cross-sections in terms of the polarizations.

For $j''=2$, 
\begin{align}
\label{OAM7}
\sigma_2={\frac{K(j''=2)}{24}}[24-5(1-pP_L) &-4(1-pP_N)\\
&-5(1-P_LP_N)], \nonumber
\end{align}
noting that the polarizations only appear in the expressions for the cross-sections two at a time. The cross-section contribution for $j''=2$ never goes to zero. It is a minimum for $pP_L=pP_N=P_LP_N=1$ and a maximum for $pP_L=P_LP_N=-1$.

For $j''=1$,
\begin{align}
\label{OAM8}
\sigma_1={\frac{K(j''=1)}{24}}[&3(1-pP_L)+(6-4\sqrt{2})(1-pP_N)\nonumber \\ 
&+(3+4\sqrt{2})(1-P_LP_N)],
\end{align}
where the irrational coefficients are the result of interference between the two possible total angular momentum states of the OAM neutron. If we assume perfect polarization of the neutron spin, OAM states, and $^3$He nuclei, $\sigma_1=0$ for $pP_L=pP_N=P_LP_N=1$ and a maximum for $pP_L=P_LP_N=-1$.

Finally, for $j''=0$, we get 
\begin{equation}
\label{OAM7}
\sigma_0={\frac{K(j''=0)}{12}}[1-pP_L+pP_N-P_LP_N].
\end{equation}

For $j''=0$, the cross-section is zero when  $pP_L=1$ or $P_LP_N=1$.  The maximum cross section is given when $pP_N=1$ and $pP_L=-1$. This is in contrast to the case for the conventional neutron singlet compound state which automatically goes to zero when $pP_N=1$ (Eq.~\ref{Rose11}). It is further seen that the presence of OAM neutrons changes the character of the singlet cross-section,  so that even if $P_L=0$, the polarization behavior is not the same. Through careful manipulation of the relative polarizations of both spins and the OAM state, the relative values of the $K(j'')$ can be determined. 

The excited-state energy levels of the $^4$He nucleus may give us an indication of the likelihood that the matrix elements $K$ are non-zero \cite{TILLEY19921}. For conventional thermal neutrons, the  interaction 
$\mathrm{n} + {}^3\mathrm{He}$ 
occurs at 20.578 MeV above the $^4$He ground state, forming a compound state with isospin $T=0$. It is conveniently 368 keV above a broad ($J^\pi = 0^+$, $T=0$) resonance at 20.21 MeV, but 7.7 MeV below the nearest ($J^\pi = 1^+$, $T=0$) energy level at 28.31 MeV. The experimental observation of neutron absorption through only the $J^\pi = 0^+$ channel is credited to the exclusive proximity of this $^4$He compound state \cite{Zimmer}. 

Owing to the $0^+$ ground state of the compound $^4$He nucleus and the addition of one unit of orbital angular momentum from the OAM neutron, we expect the final states of the compound nucleus to be odd parity, $J^\pi=0^-$, $1^-$, and $2^-$.
An OAM neutron absorbed by $^3$He would form a compound nucleus in the immediate vicinity of  broad excited states of $^4$He ($J^\pi = 0^-$, $T=0$) lying at 21.01 MeV and ($J^\pi = 2^-$, $T=0$) at 21.84 MeV, respectively, while the nearest level ($J^\pi = 1^-$, $T=0$) lies considerably higher at 24.25 MeV. The decay products of these levels may differ from  Eq.~\ref{Interaction} although available data is based on different interactions \cite{TILLEY19921}.
In particular, while the $0^+$ excited state decays by emitting a proton, both the aforementioned $0^-$ and $2^-$ states can decay by reemitting a neutron.

In conclusion, we propose a detection method for individual neutrons put into intrinsic orbital angular momentum states that depends on an unambiguous quantum effect, the nuclear interaction with $^3$He. We have shown that when a thermal, spin-polarized neutron, put into a helical $L=1$ OAM state with specific $m_L$ along its wavevector, is absorbed by spin-polarized $^3$He, the compound nuclear state should be readily distinguished from the case of an ordinary neutron. Three possible final states may result, with a total angular momentum  $J^\pi=0^-$, $1^-$, and $2^-$. The dependence of the possible absorption cross-sections on the polarizations of the initial states will differ considerably from the dependence for ordinary neutrons.

As in the case for ordinary neutrons, there are energetic grounds for assumption that the compound nucleus may not form in all three total angular momentum channels. Furthermore, because the parity of the final state is negative, we have reason to believe that the nuclear interaction part of the matrix element will be distinct from the positive parity case of the singlet or triplet in the compound state formed with ordinary neutrons. The creation in quantity of thermal OAM neutrons offers potential for new physics, including alternative channels for radioactive decay, modified cross sections for nuclear interactions, and previousy unobserved nuclear processes.

\begin{acknowledgments}
The authors acknowledge useful discussions with R. Cappelletti and C. Majkrzak.
\end{acknowledgments}

\bibliography{He3_OAM}
\end{document}